# Characterization of NbTiN/HZO/NbTiN MIM Capacitors for high frequency AC Clock & Power Distribution Network for Superconducting Digital Circuits

Seifallah Ibrahim[1*], Blake Hodges[1], Steven Brebels[2], Julian Gil Pinzon[1], Trent Josephsen[1], Ankit Pokhrel[2], Daniel Perez Lozano[2], Yann Canvel[2], Bart Kenens[2], Amey M. Walke[2], Gianpiero Maccarrone Lapi[1], Sara Iraci[1], Mihaela Popovici[2], Benjamin Huet[2], Quentin Herr[1,2], Zsolt Tőkei[2], and Anna Herr[1,2]

[1]imec USA, Kissimmee, FL, USA; [2]imec, Heverlee, Belgium.
*Corresponding email: seifallah.ibrahim@imec.be

**Abstract** – A resonant clock-power distribution network is critical for scaling energy efficient superconducting digital technology to practical high integration density circuits. High-k, tunable capacitors enable implementation of a resonant power delivery network supporting circuits with up to 400 Mdevices/cm$^2$. We report the cryogenic characterization of Metal-Insulator-Metal capacitors using a Hafnium Zirconium Oxide (HZO) ferroelectric insulating layer and Niobium Titanium Nitride (NbTiN) superconducting electrodes. The fabricated chip includes capacitor arrays for low frequency characterization and a half-wave transmission line resonator for RF characterization. A specific capacitance of 3 $\mu F/cm^2$, DC leakage current of $10^{-8}$ A/cm$^2$ at 2 V and a constant 5% tunability up to 4 GHz were measured at 2.6 K.

## Introduction

Demand for compute power and hardware in the post-Moore's Law era is outpacing hardware capabilities [1], opening the door for transformative technologies. Superconducting digital (SCD) technology has the potential to revolutionize high performance compute and AI with energy-efficient active devices, high computational density, and high bandwidth interconnect, providing a sustainable path beyond CMOS [2]. SCD technology energy efficiency comes from the fundamental physics of superconducting materials providing zero resistance wires and low energy switching Josephson junctions (JJs). However, energy efficient and scalable power delivery has been a major challenge [3]. The technology has evolved from DC power distribution using resistors to a resonant AC power distribution network (PDN). The first PDNs [4] showed static power on par with JJ power and ability to deliver uniform current to the JJ in series across 1 cm die at 5 GHz. Scaling of the first PDNs in frequency and density has been limited by magnetic transformers. Imec has developed a scalable approach for the AC resonant PDN based on a 2D array of coupled LC resonators (Fig. 1). Use of capacitive coupling to the JJ enables clock frequency of 30 GHz and density of bias taps of 400 Mtaps/cm$^2$ with a PDN efficiency of 90% for a fully utilized circuit (Fig. 2). The high-k tunable Hafnium Zirconium Oxide (HZO) Metal-Insulator-Metal (MIM) capacitor with superconducting Niobium Titanium Nitride (NbTiN) electrodes is the key new element of the AC PDN [5]. The superconducting PDN is high-Q, so tunability is needed to compensate for fabrication induced resonance frequency mistargeting. The non-linearity of the HZO-based ferroelectric (FE) MIM capacitor [6] makes it an ideal candidate for a voltage-controlled tunable resonator. A reference model for the relevant parts of the fabrication stack (Fig. 3) shows that the capacitive PDN can be implemented using the same processes developed at imec for NbTiN wire up [7] and is compatible in density with the new high critical current density amorphous-Si JJ devices [5].

Here we report performance data for NbTiN/HZO/NbTiN capacitors fabricated on a 300 mm wafer with critical dimensions (CDs) down to 190 nm. For the first time, these capacitors were characterized at cryogenic (cryo) temperatures, down to 2.6 K, in terms of PDN critical metrics: leakage current, the effects of thermal and electrical cycling on FE properties, and the tunability from low frequency to 4 GHz.

## Low frequency characterization

The low frequency (LF) test structure shown in Fig. 4.b consists of an array of 9.5 nm thick HZO capacitors connected in parallel by top and bottom 50 nm thick NbTiN electrodes fabricated as described in [5]. This array was used to characterize the ferroelectric performance across the following parameters: 1) *Coercive Voltage ($V_c$)*, the minimum voltage required to switch polarity of the ferroelectric domains, 2) *Peak Capacitance ($C_{peak}$)*, the maximum capacitance obtained at $V_c$, 3) *Tunability (%)*, the total percent change in capacitance as a function of applied DC bias voltage, and 4) *Saturation Voltage ($V_s$)*, the voltage at which ferroelectric domain switching is effectively complete $(dQ/dV \sim 0)$. The same structure was used to characterize the LF *leakage current*.

Fig. 5 shows room temperature (RT) C-V measurements of the capacitor array test structures. Measurements performed across two wafers and 30 structures demonstrated >90% yield, with specific capacitance of $C_f \sim 3$ $\mu F/cm^2$ and tunability of up to 10%. The measurements showed correlation between properties of the capacitors and properties of the NbTiN film with known concentric variation cross-wafer [8]. As shown in Fig. 5, wafer-edge devices with the desired NbTiN film properties have symmetric coercive and saturation voltages of $V_c = \pm 1$ V and $V_s = \pm 2$ V. Peak capacitance of $C_{peak}(V_c) = 320$ pF (3.3 $\mu F/cm^2$) measured across 7 devices sampled from 4 edge locations. At a fixed bias of 0.9 V, the capacitance spread is below 1% across wafer locations and below 0.2% across the die.

Repeating the measurements at 2.6 K (Fig. 6) shows a reduction in both tunability (~5%) and nominal capacitance (2.9 $\mu F/cm^2$), and an increase in $V_c$ of less than 0.1 V. The temperature dependence on both capacitor tunability and $V_c$ is attributed to the lack of thermal excitation energy limiting the mobility of the ferroelectric domains at cryogenic temperatures. Both tunability and the permittivity of the HZO capacitors show a linear dependence with temperature down to 7 K (Fig. 6.b). The parameters of capacitors remain constant across the operating temeapartures of SCD circuits of 2.6-5 K.

The positive-up negative-down (PUND) technique [9] was used to decouple the FE switching current from leakage, capacitive displacement, and charging/discharging current components. The results (Fig. 7) show that capacitors are symmetric regarding ferroelectric-domain switching behavior with FE switching current $|I_{sw+}| \sim |I_{sw-}| = 1$ nA $\pm 0.1$.

A second series of tests accurately isolated leakage current by applying a voltage step after fully discharging the capacitor and measuring the current response over the long settling time. The leakage current of one capacitor array is shown in Fig. 8 at RT and 2.6 K. Low leakage current density of $J(2V) = 3\text{x}10^{-5}$ A/cm$^2$ was measured at RT both inline during fabrication and confirmed in the cryogenic test lab. Cryo test showed a three-orders-of-magnitude decrease in leakage current down to $J(2V) = 4.5\text{x}10^{-8}$ A/cm$^2$. The measured leakage current is well below the application requirement of 1 $\mu$A per $1 \times 1$ cm chip with 400 Mtaps.

## High frequency characterization

High frequency (HF) characterization of the capacitors used a half-wave (HW) transmission line (TL) resonator (Fig. 9). The HW resonator is formed between top (signal) and bottom (ground) NbTiN electrodes connected via the HZO capacitors. The transmission and capacitor geometry are chosen to target a 50 Ω characteristic impedance. The resonator's scattering parameters are measured up to 10 GHz with applied DC bias of 0-3 V as shown in Fig 10. The capacitor tunability is extracted from the resonance frequency shift. The 5% tunability is maintained up to at least 4 GHz. Measurement at yet higher frequencies was limited by the lossy-path through the highly-doped Si substrate. However, the doped substrate did allow for room-temperature in-line characterization (probe-to-chuck C-V).

## Conclusion

We characterized 9.5 nm thick HZO MIM capacitors with NbTiN electrodes using a capacitor array for low frequency test and a resonator for high frequency test. As summarized in Table 1, this work pioneers characterization of HZO capacitors with 190 nm linear dimension, cryo-test down to 2.6 K, and HF test up to 4 GHz. The devices exhibit the desired electrical performance in terms of specific capacitance, tunability, and saturation voltage at cryo temperature. The leakage current at cryo is extremely small. The properties of the capacitors at HF do not degrade up to 4 GHz. These results show applicability of HZO MIM capacitors for implementation of a high-density, energy efficient resonant power network for superconducting digital circuits.

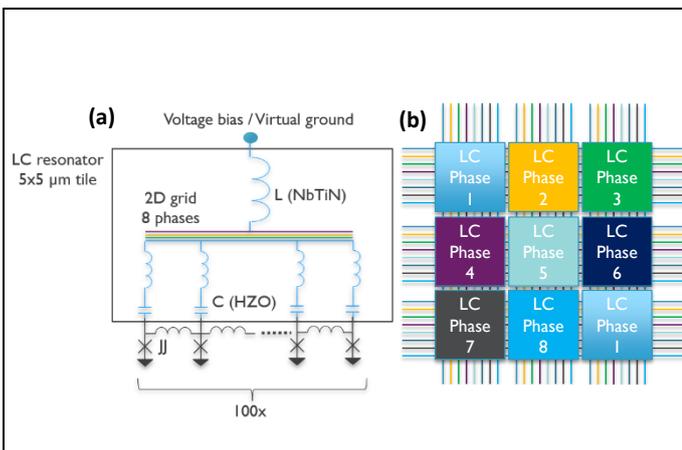

Fig 1. (a) Conceptual schematic of LC resonant PDN to Josephson junctions consisting of NbTiN inductors and parallel capacitor arrays. Resonator inductors of opposite phase are connected to create a virtual ground. This configuration enables the application of a DC voltage bias to tune HZO MIM capacitors, allowing tunability of the resonant frequency. (b) Block diagram of a 2D chip level resonant PDN with multiple phases providing clock and phase uniformity across 1 cm chip at 30GHz.

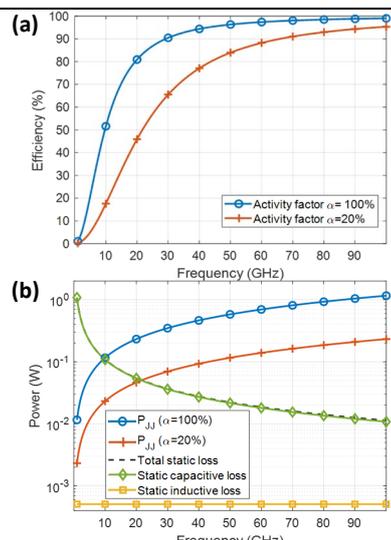

Fig 2. Modeled performance of the resonant PDN based on HZO MIM capacitors. (a) Room temperature equivalent power consumption per JJ including 325W/W cryocooler efficiency, (b) PDN efficiency relative to JJ activity factor in a circuit.

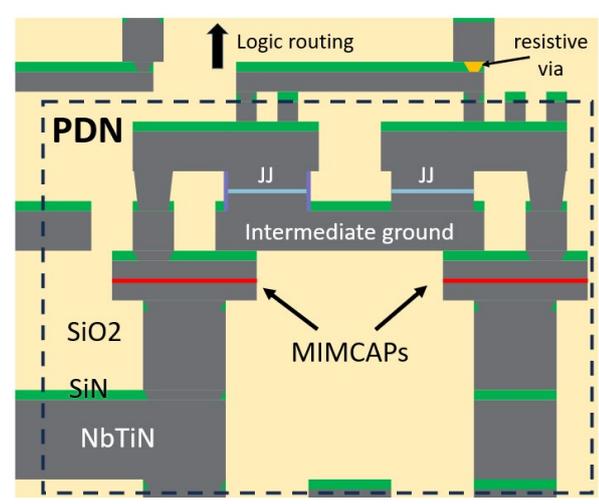

Fig 3. Cross-section of a portion of the proposed fabrication stack consisting of NbTiN/a-Si/NbTiN Josephson junctions with NbTiN/HZO/NbTiN MIM capacitors for the power network. The structures characterized in this paper are based on the MIM cap module.

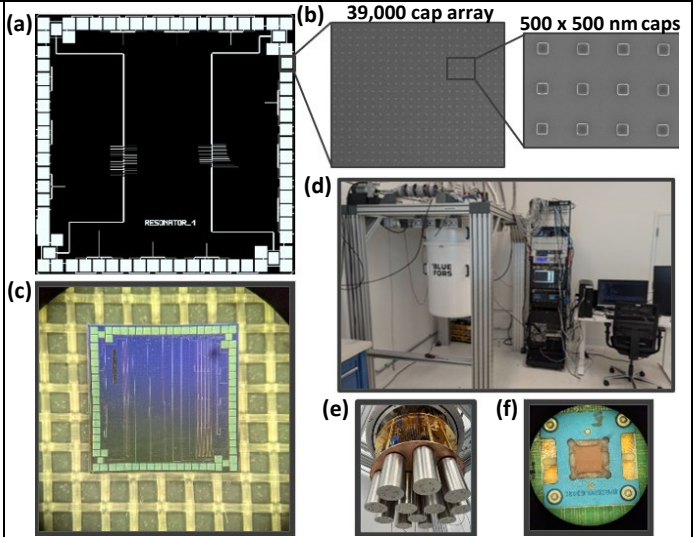

Fig 4. DC test structures shown in (a) layout and (b) top-down SEM images of capacitor arrays populating contact pads. Images of (c) bare die, (d) cryogenic test lab with BlueFors XLDsl-4K cryocooler sealed, and (e) open. (f) Device-interface board with spring-loaded pogo-pins rated up to 40 GHz.

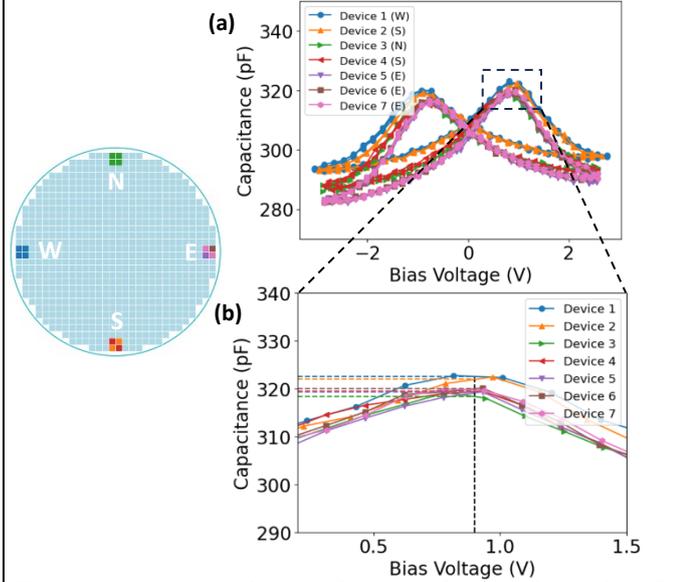

Fig 5. (a) RT CV hysteresis loops of 7 devices from 4 wafer locations. (b) Coercive voltage spread <0.1V, capacitance spread <1% at fixed bias of 0.9V

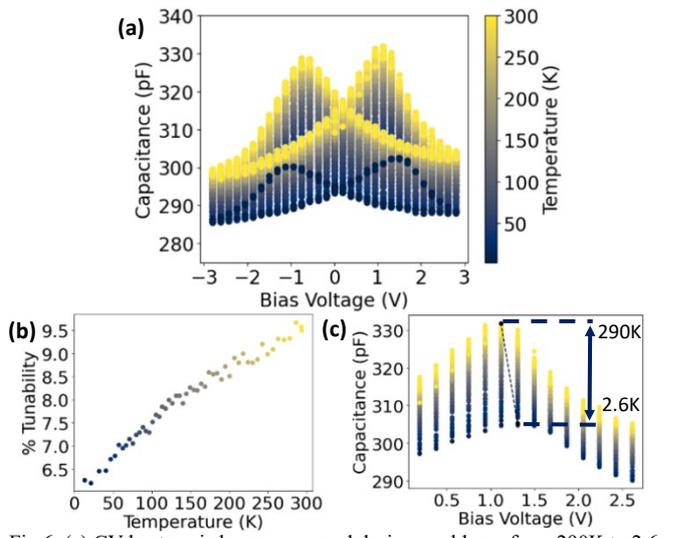

Fig 6. (a) CV hysteresis loops measured during cooldown from 290K to 2.6 K. (b) Tunability-temperature dependence down to 2.6 K, linear decrease in tunability. 10% RT – 5% 2.6 K and (c) <0.1V increase in coercive voltage from RT to 2.6K ($V_c$ accuracy limited by measurement resolution)

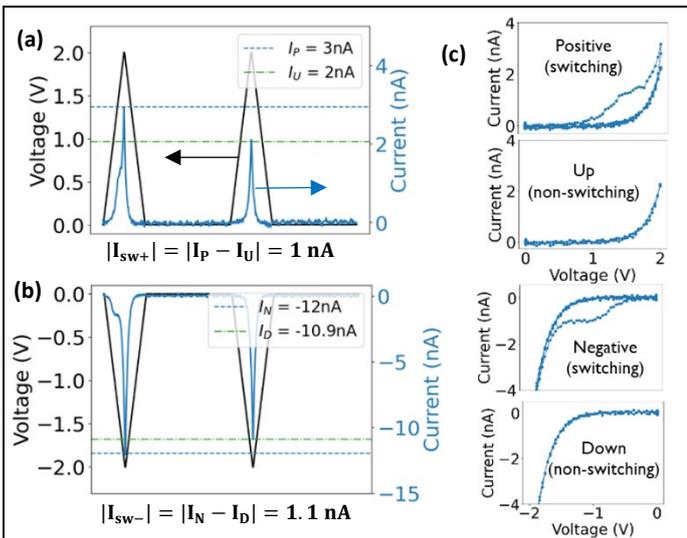

Fig 7. PUND measurement demonstrating FE symmetry and showing FE switching only on the first pulse of either polarity as expected. Time-domain applied voltage and measured current for (a) Positive-Up and (b) Negative-Down pulses, and (c) corresponding IV curves for each pulse.

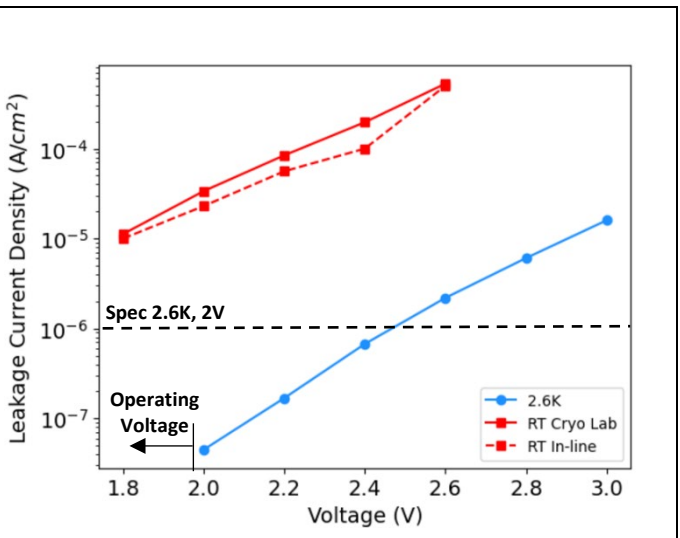

Fig 8. Leakage current as function of applied DC voltage at room temperature and at cryo, showing 1.5 orders-of-magnitude lower cryo leakage current at 2 V than the target specification.

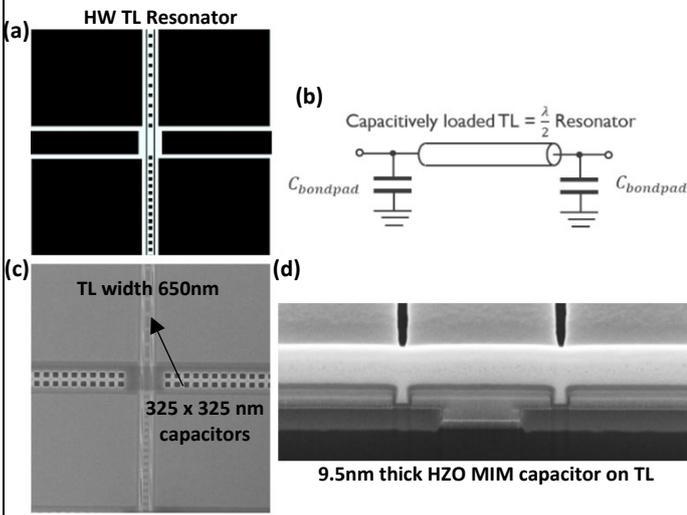

Fig 9. (a) Layout and (b) schematic of half-wave transmission line (TL) resonator fabricated on the same chip as in Fig. 4, used to measure high frequency FE response of HZO capacitors. (c) Top-down SEM image and (d) cross-section TEM of resonator unit cell.

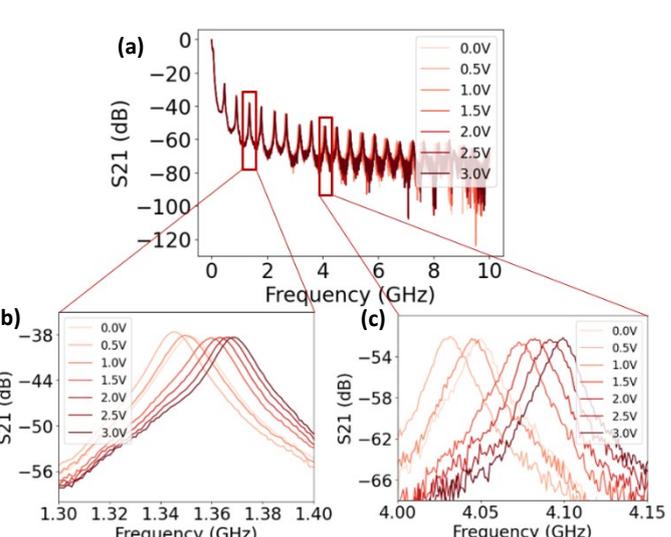

Fig 10. (a) S21 parameter of the TL resonator shown with applied DC bias across HZO capacitors. Resonance frequency tunability of 5% measured (b) at RF frequency around 1.36 GHz, and (c) and again around 4.07 GHz.

Table I. Comparison of cryogenic & HF HZO capacitor data.

|  | EDS'22 [10] | SC'24 [11] | DRC'21 [12] | This work |
|---|---|---|---|---|
| Experiment | Cryo | Cryo | HF | Cryo + HF |
| HZO Thickness | 6.3 nm | 10 nm | 10 nm | 9.5 nm |
| Minimum Temperature | 77 K | 30 K | 293 K (RT) | 2.6 K |
| $(V_{c-}, V_{c+})$ (V) | (-0.2, 1) Non-symmetric | (-1.1, 0.8) Non-symmetric | (-1.1, 1.5) Non-symmetric | (-1.1, 1.1) Symmetric |
| Tunability (+ve branch) | ~5.5% @77K | ~10% @30K | ~4.5% @RT Degrades w/ freq | 5% @2.6K Non-degrading |
| Frequency | LF | LF | 2 GHz | >4 GHz |
| Leakage (2V) (A/cm$^2$) |  | 6x10$^{-7}$ |  | 4.5x10$^{-8}$ |

**Acknowledgements** Work is funded in part by imec INVEST+ fund, Osceola County FL, NSF Central Florida Semiconductor Engine, and AFRL contract W911NF-24-1-0150.